\documentclass[letterpaper, 10 pt, conference]{ieeeconf}
\IEEEoverridecommandlockouts
\usepackage{algorithm} 

\usepackage{amsthm}
\usepackage{caption}
\usepackage{float} 


\usepackage[normalem]{ulem}
\usepackage{cancel}
\usepackage[dvipsnames]{xcolor}
\usepackage{epsfig,epsf,epstopdf}
\graphicspath{ {./Figs/} }
\usepackage[cmex10]{amsmath}
\usepackage{amsfonts}
\usepackage{amssymb}
\usepackage[noend]{algpseudocode}
\makeatletter
\def\BState{\State\hskip-\ALG@thistlm}
\makeatother
\usepackage{cite}
\usepackage{paralist}
\usepackage{color}
\usepackage{array}
\usepackage{mathrsfs}
\usepackage{multirow}
\usepackage{subfigure}
\usepackage{graphicx}
\usepackage{caption}
\usepackage[font=small,justification=justified,singlelinecheck=false]{caption}

\usepackage{nomencl}

\newtheoremstyle{boldhead}
  {1pt} 
  {1pt} 
  {\normalfont} 
  {} 
  {\bfseries\itshape} 
  {.} 
  { } 
  {\thmname{#1}\thmnumber{ #2}\thmnote{ \normalfont(#3)}} 

\theoremstyle{bolditalichead}
\newtheorem{theorem}{Theorem}

\newtheorem{assumption}{Assumption}
\newtheorem{definition}{Definition}
\newtheorem{remark}{Remark}
\newtheorem{lemma}{Lemma}

\definecolor{blue}{rgb}{0.0, 0.0, 0.65}
\usepackage{hyperref}

\makenomenclature
\begin{document}
\title{Certified Resilient Collision-free Control Framework for Heterogeneous Multi-agent Systems Against Exponentially Unbounded FDI Attacks}

\author{Yichao Wang, Mohamadamin Rajabinezhad, Dimitra Panagou, and Shan Zuo
\thanks{Yichao Wang, Mohamadamin Rajabinezhad, and Shan Zuo are with the Department of Electrical and Computer Engineering, University of Connecticut, Storrs, CT 06269, USA. Dimitra Panagou is with the Department of Robotics and the Department of Aerospace Engineering, University of Michigan, MI 48109, USA. (Emails: yichao.wang@uconn.edu; mohamadamin.rajabinezhad@uconn.edu; dpanagou@umich.edu; shan.zuo@uconn.edu.)}
}


\maketitle

\thispagestyle{empty} 

\begin{abstract}
False data injection attacks pose a significant threat to autonomous multi-agent systems (MASs). Existing attack-resilient control strategies generally have strict assumptions on the attack signals and overlook safety constraints, such as collision avoidance. In practical applications, leader agents equipped with advanced sensors or weaponry span a safe region to guide heterogeneous follower agents, ensuring coordinated operations while addressing collision avoidance to prevent financial losses and mission failures. This letter addresses these gaps by introducing and solving the safety-aware and attack-resilient (SAAR) control problem under exponentially unbounded false data injection (EU-FDI) attacks. Specifically, a novel attack-resilient observer layer (OL) is first designed to defend against EU-FDI attacks on the OL. Then, an attack-resilient compensational signal is designed to mitigate the adverse effects caused by the EU-FDI attack on control input layer (CIL). Finally, a SAAR controller is designed by solving a quadratic programming (QP) problem integrating control barrier function (CBF) certified collision-free safety constraints. Rigorous Lyapunov-based stability analysis certifies the SAAR controller's effectiveness in ensuring both safety and resilience. This study also pioneers a three-dimensional (3D) simulation of the SAAR containment control problem for heterogeneous MASs, demonstrating its applicability in realistic multi-agent scenarios.
\end{abstract}


\section{Introduction}
Containment control in autonomous MASs has become essential in real-world applications, particularly involving unmanned aerial vehicles (UAVs) \cite{petrlik2020robust} and unmanned ground vehicles (UGVs) \cite{he2020ground}. This control strategy enables leader agents, often equipped with advanced sensors or weaponry, to guide follower agents within a designated safe region, ensuring coordinated movement and operational safety. Such scenarios are common in tasks such as surveillance, reconnaissance, and military operations, where maintaining cohesion and mitigating risks are critical. The heterogeneity of agents, characterized by differing models, structures, or capabilities, adds complexity to the problem, necessitating advanced control algorithms. Moreover, collision avoidance plays a pivotal role in containment control, as collisions can lead to substantial financial losses and mission failures, highlighting the importance of integrating safety measures into control strategies.

In the MAS framework, FDI attacks pose significant threats to autonomous MASs and impact critical infrastructures and mission-essential cyber-physical systems, such as power systems, transportation networks, and combat-zone multi-robot systems \cite{weng2023secure}. To mitigate the impact of FDI attacks, two primary approaches have been proposed. The first one involves detection and identification of compromised agents, followed by the attack-isolation \cite{zhang2021co,pasqualetti2013attack}. However, this method relies on strict limitations, such as the number of compromised agents, which are often impractical. To address these limitations, a second approach has been developed, focusing on the design of attack-resilient control protocols \cite{zuo2022resilient,zuo2023resilient,shi2024resilient,liao2024resilient,zhang2024resilient}. Rather than detecting and removing compromised agents, this strategy aims to minimize the adverse effects of attacks through resilient control mechanisms. However, the aforementioned research neglects safety constraints, such as collision avoidance, leading to potential collision risks. On the other hand, existing safe control methods incorporating safety constraints often fail to guarantee resilience for MASs against FDI attacks \cite{garg2022fixed,cohen2024constructive,wu2023quadratic,cohen2023characterizing,tan2024undesired}. Additionally, most studies addressing FDI resilience for heterogeneous MASs assume the OL remains intact and either disregard attacks or handle only bounded attack signals or the ones with bounded first-time derivatives, which are impractical assumptions \cite{ye2024decentralized,zuo2023resilient}. However, malicious adversaries could inject \textit{any time-varying} signals, such as EU-FDI attacks, to compromise the system.

In real-world scenarios, ensuring both safety and resilience against EU-FDI attacks is crucial for the reliable operation of safety-critical MASs. To address this challenge, this letter formulates and solves the SAAR control problem. The key difficulty lies in designing a controller that effectively mitigates EU-FDI attacks while consistently enforcing collision avoidance constraints. The main contributions are as follows.\\
$\bullet$ A novel attack-resilient OL is first designed to defend against EU-FDI attacks on OL. Then, an attack-resilient compensation signal is developed to mitigate the adverse effects of EU-FDI attacks on CIL. Finally, the SAAR controller is designed by solving a QP problem incorporating CBF-certified collision-free safety constraints. To the best of the authors' knowledge, this letter is the first to guarantee both resilience and safety for heterogeneous MASs, in the presence of EU-FDI attacks on both OL and CIL.\\
$\bullet$ A rigorous Lyapunov-based stability analysis is conducted to certify the effectiveness of the proposed SAAR controller in ensuring both safety and resilience. Its efficacy is further demonstrated through a 3D simulation of the containment control problem for heterogeneous MASs.

\vspace{-2mm}
\section{Preliminaries and Problem Formulation}
\( \|x\| \) denotes the Euclidean norm of a vector \( x \in \mathbb{R}^n \).

\noindent \( L_f V(x)=\frac{\partial V}{\partial x} f(x) \) is the Lie derivative of a function \( V : \mathbb{R}^n \rightarrow \mathbb{R} \) along a vector field \( f : \mathbb{R}^n \rightarrow \mathbb{R}^n \). \( {I_N} \in {\mathbb{R}^{N \times N}} \) is the identity matrix. \( {\mathbf{1}}_N, {\mathbf{0}}_N \in {\mathbb{R}^N} \)  are the column vectors with all elements of zero and one, respectively. The Kronecker product is represented by \( \otimes \). The operator \( \operatorname{diag}(\cdot) \) forms a block diagonal matrix from its argument. The notations \( {\sigma_{\min}}(\cdot) \) and \( {\sigma _{\max }}(\cdot) \) represent the minimum singular value and the maximum singular value respectively. Consider a system of \( N + M \) agents on a time-invariant digraph \( \mathscr{G} \), consisting of \( N \) followers and \( M \) leaders, denoted by \( \mathscr{F} = \{v_1, v_2, \ldots, v_N\} \) and \( \mathscr{L} = \{v_{N+1}, v_{N+2}, \ldots, v_{N+M}\} \), respectively. \( \mathcal{N}_i \) represents the set of neighboring followers of follower \( i \). The interactions among followers are described by the subgraph \( \mathscr{G}_f = (\mathcal{V}, \mathcal{E}, \mathcal{A}) \). More details on the digraph can be found in \cite{zuo2022resilient}.
Denote \(\Phi_r = \frac{1}{M}{\mathcal{L}} + {\mathcal{G}}_r\). The states of the leaders are represented by the set \( X_{\mathscr{L}} = \{x_{N+1}, x_{N+2}, \ldots, x_{N+M}\} \). 

We consider a group of \( N \) heterogeneous followers described by the following dynamics:
\begin{equation}
\dot{x}_i(t) = A_i x_i(t) + B_i u_i(t), \quad i \in \mathscr{F},
\label{eq: follower dynamics}
\end{equation}
where \( x_i(t) \in \mathbb{R}^{n} \) is the state, \( u_i(t) \in \mathbb{R}^{m_i} \) is the control input. The system matrices $A_i$ and $B_i$ may vary across different agents, making the system heterogeneous. The $M$ leaders with the following dynamics can be viewed as command generators that generate the desired trajectories:
\begin{equation}
\label{eq: leader dynamics}
\dot{x}_r(t) = S x_r(t), \quad r \in \mathscr{L},
\end{equation}
where $x_r(t) \in \mathbb{R}^{n}$ is the state of the $r$th leader.

\begin{definition}[\cite{rockafellar2015convex}]
\label{def: convex_hull}
The convex hull \( \text{Co}(X_{\mathscr{L}}) \) is the minimal convex set containing all points in \( X_{\mathscr{L}} \), defined as:
\[
\text{Co}(X_{\mathscr{L}}) = \left\{ \sum\nolimits_{r\in \mathscr{L}} a_r x_r \;\middle|\; a_r \geqslant 0, \sum\nolimits_{r\in \mathscr{L}} a_r = 1 \right\}.
\]
where $\sum_{r\in \mathscr{L}} a_r x_r$ represents certain convex combination of all points in $X_{\mathscr{L}}$.
\end{definition}

\begin{definition}[\cite{khalil2002nonlinear}]
\label{def: UUB}
The signal $x(t)\in {\mathbb{R}^n}$ is said to be UUB with the ultimate bound $b$, if there exist positive constants $b$ and $c$, independent of ${t_0} \geqslant 0$, and for every $a \in \left( {0,c} \right)$, $\exists \, T = T\left( {a,b} \right) \geqslant 0$, independent of $t_0$, such that
\begin{equation}
\label{eq12}
\left\| {x\left( {{t_0}} \right)} \right\| \leqslant a\;\; \Rightarrow \;\;\left\| {x\left( t \right)} \right\| \leqslant b,\forall t \geqslant {t_0} + T
\end{equation}
\end{definition}

\begin{assumption}
\label{ass: leader follower}
Each follower in the digraph ${\mathscr{G}}$, has a directed path from at least one leader.
\end{assumption}

\begin{assumption}
\label{ass: eig of S}
$S$ has eigenvalues with non-positive real parts, and non-repeated eigenvalues on the imaginary axis.
\end{assumption}

\begin{assumption}
\label{ass: controllable A and B}
The pair $(A_i,B_i)$ is controllable for each follower.
\end{assumption}

Given Assumption~\ref{ass: controllable A and B}, the following linear matrix equation has a solution \( \Pi_i \) for each follower:
\begin{equation}
S = A_i + B_i \Pi_i.
\label{eq: solution for tracking}
\end{equation}

\begin{assumption}
\label{ass: attacks} The FDI attack signals on the CIL and OL,
 $\gamma^a_i(t)$ and $\gamma^{ol}_i(t)$, are exponentially unbounded. That is, their norms grow at most exponentially with time. For the purposes of stability analysis, it is reasonable to assume that there exist positive constants $\kappa^a_i$ and $\kappa^{ol}_i$, such that
$\|\gamma^a_i(t)\| \leqslant \exp(\kappa^a_i t)$ and $ \|\gamma^{ol}_i(t)\| \leqslant \exp(\kappa^{ol}_it)$, where $\kappa^a_i$ and $\kappa^{ol}_i$ could be unknown.
\end{assumption}

\begin{remark}
\label{rem: remark on the attack signals}
Assumption \ref{ass: attacks} covers a broad range of FDI attack signals, including those that grow exponentially over time. Specifically, \( \exp(\kappa^a_it) \) and \( \exp(\kappa^{ol}_it) \) represent the worst-case scenarios the controller can handle. As long as the attack signals remain below these envelops, the controller can effectively mitigate them. In practice, adversaries can inject \textit{any time-varying signal} into MAS through platforms such as software, CPUs, or DSPs. However, most of the existing research focuses on disturbances, noise, or bounded attack signals, or assumes that the first time derivatives of the attacks are bounded \cite{zuo2023resilient}.
\end{remark}

Let $\bar x_r= \mathbf{1}_N\otimes x_r$. Define the global containment error as
\begin{equation}
\begin{gathered}
e_c = x - \left( \sum\nolimits_{\nu \in \mathscr{L}} \left( \Phi_\nu \otimes I_n \right) \right)^{-1} \sum\nolimits_{r \in \mathscr{L}} \left( \Phi_r \otimes I_n \right)\bar x_r,
\end{gathered}
\label{eq: containment error vector}
\end{equation}
where $x = {[ {x_{1}^\top,...,x_{N}^\top} ]^\top}$.
\begin{lemma}[\cite{zuo2017output}]
\label{le: covex combinations}Given Assumption~\ref{ass: leader follower}, the containment control objective is achieved if \( \lim_{t \to \infty} e_c(t) = \mathbf{0} \). That is, each follower converges to the convex hull spanned by the leaders.
\end{lemma}


\begin{definition}
\label{def: invariant set}
A set \( S \subset \mathbb{R}^n \) is forward invariant with respect to a control system if, for all initial conditions \( x(0) \in S \), the system's trajectory satisfies \( x(t) \in S \) for all \( t \geqslant 0 \).
\end{definition}

The forward invariance of the safe set is guaranteed using CBFs, as described as follows.

\begin{definition}[\cite{ames2014control}]
\label{def: general CBF}
Let \( \alpha \) be a monotonically increasing, locally Lipschitz class \( \mathcal{K} \) function with \( \alpha(0) = 0 \), given a set \( S \subset \mathbb{R}^n \) as defined in Definition~\ref{def: invariant set}, a function \( h(x) \) is CBF, if the following condition is satisfied:
\begin{equation}
\inf_{u \in \mathcal{U}} \left[ L_{f_i} h(x) + L_{g_i} h(x) u \right] \leqslant -\alpha(h(x)),
\label{eq: general CBF condition}
\end{equation}
where \( \mathcal{U} \) is the set of admissible control inputs.
\end{definition}

In \eqref{eq: general CBF condition}, \( L_f h(x) \) and \( L_g h(x) \) denote the Lie derivatives of the function \( h(x) \) along the vector fields \( f \) and \( g \), respectively. For a general system of the form: \(\dot{x} = f(x) + g(x)u,
\)
the terms \( L_f h(x) = \frac{\partial h}{\partial x} f(x) \) and \( L_g h(x) = \frac{\partial h}{\partial x} g(x) \) involve the gradients of \( h(x) \) with respect to the state \( x \). if control input $u$ is designed such that \eqref{eq: general CBF condition} is satisfied, the forward invariant set $S$ is obtained \cite{wang2017safety}.

Now, consider our specific case where \(h(x)\) applied locally corresponds to \( h_{S_{ij}}(x_i, x_j) \). We apply aforementioned general concepts to our specific case, where the safe set for collision avoidance is defined as:
\begin{equation}
S_{S_{ij}} \triangleq \{x_i \in \mathbb{R}^n \mid h_{S_{ij}}(x_i, x_j) \leqslant 0\},
\label{eq: safe set}
\end{equation}
with the specific CBF:
\begin{equation}
h_{S_{ij}}(x_i, x_j) \triangleq d_s^2 - \|x_i - x_j\|^2,
\label{eq: CBF}
\end{equation}
where \( d_s \) is the minimum allowable safe distance. The system dynamics are given by: \(
\dot{x}_i(t) = A_i x_i(t) + B_i u_i(t), \, \forall i \in \mathscr{F}.
\)
In this case, the vector field \( f_i(x) \), applied to linear time invariant system locally, corresponds to \( A_i x_i \), and \( g_i(x) \) corresponds to \( B_i\). To avoid confusion, we denote the Lie derivatives as \( L_{f_i} h_{S_{ij}} \) and \( L_{g_i} h_{S_{ij}} \), where \( f_i \) and \( g_i \) are defined as above.

The following definition introduces the SAAR control problem.

\begin{definition}[SAAR Control Problem]
\label{def: Safety-aware and Attack-Resilient Control Problem}
For the heterogeneous MAS described in \eqref{eq: follower dynamics}-\eqref{eq: leader dynamics} under EU-FDI attacks, the SAAR control problem is to design control input $u_i \in \mathcal{U}_i$, such that 1) $e_c$ is UUB, i.e., the UUB containment control objective is achieved. Specifically, the state of each follower converges to a small neighborhood around or within the dynamic convex hull spanned by the states of the leaders; 2) the state \( x_i(t) \) remains within the safe set \( S_{S_{ij}} \) for all \( t \geqslant 0 \) in the presence of EU-FDI attacks on both OL and CIL.
\end{definition}
\section{Main Result of SAAR Controller Design}
In this section, we propose a fully distributed, safe, and resilient containment control framework to address the SAAR control problem. Since only the neighboring followers of the leaders have access to the leaders' states, to achieve the containment control objective for each follower, a OL dynamics is needed to estimate the convex combinations of the leaders' states. Here, we introduce the following fully distributed and attack-resilient OL dynamics
\begin{equation}
\dot{\zeta}_i = S \zeta_i + \exp(\vartheta_i) \xi_i + \gamma^{ol}_i,
\label{eq: single agent observer dynamics}
\end{equation}
\begin{equation}
\dot{\vartheta}_i = q_i \xi_i^\top \xi_i,
\label{eq: adaptive gain theta_i dynamics}
\end{equation}
where $\zeta_i$ is the local state on the OL and \( \gamma^{ol}_i \) represents the attack signal on the OL, $\vartheta_i$ is adaptively tuned by \eqref{eq: adaptive gain theta_i dynamics} with constant \( q_i > 0 \), and $\xi_i$ represents the gathered neighborhood relative information on the OL given by
\begin{equation}
\xi_i = \sum_{j \in \mathscr{F}} a_{ij} (\zeta_j - \zeta_i) + \sum_{r \in \mathscr{L}} g_{ir} (x_r - \zeta_i), i \in \mathscr{F}
\label{eq: containment information}
\end{equation}

Based on this OL, we then introduce the following conventional control input design \cite{zuo2017output}.
\begin{equation}
\begin{gathered}
  u^c_i = K_i x_i + H_i \zeta_i, \hfill 
\end{gathered}
\label{eq: conventional control input}
\end{equation}
The matrices \( K_i \) and \( H_i \) are obtained as:
\begin{equation}
K_i = - U_i^{-1} B_i^\top P_i,
\label{eq: K_i equation}
\end{equation}
\begin{equation}
H_i = \Pi_i - K_i,
\label{eq: H_i equation}
\end{equation}
where the matrix \( P_i \) is obtained by solving the following algebraic Riccati equation\cite{zuo2023resilient}:
\begin{equation}
A_i^\top P_i + P_i A_i + Q_i - P_i B_i U_i^{-1} B_i^\top P_i = 0.
\label{eq: Riccati equation}
\end{equation}

Define the follower-observer tracking error as
\begin{equation}
\varepsilon_i = x_i - \zeta_i.
\label{eq: follower-observer error vector}
\end{equation}

We consider the additional EU-FDI attack signal in the CIL for each follower, $\gamma^a_i$. To address these attacks, we then design the following attack-resilient control input signal
\begin{align}
u^r_i &= u^c_i - \hat{\gamma}^a_i,
\label{eq: control input}\\
\hat{\gamma}^a_i &= \frac{B_i^\top P_i \varepsilon_i}{\| \varepsilon_i^\top P_i B_i \| + \exp(-c_i t^2)} \exp(\hat{\rho}_i),
\label{eq: compensational signal}\\
\dot{\hat{\rho}}_i &= \alpha_i \| \varepsilon_i^\top P_i B_i \|,
\label{eq: adaptively tuned parameter}
\end{align}
where the compensational signal \( \hat{\gamma}^a_i \) mitigates the adverse effects caused by the attack $\gamma^a_i$, and \( \hat{\rho}_i \) is adaptively tuned by \eqref{eq: adaptively tuned parameter} with constant $\alpha_i > 0$. Now the corrupted attack-resilient controller becomes $\bar u_i= u^r_i + \gamma^a_i$.

Define the observer containment error vector as
\begin{equation}
\begin{gathered}
\Delta_o = \zeta - \left( \sum_{\nu \in \mathscr{L}} \left( \Phi_\nu \otimes I_n \right) \right)^{-1} \sum_{r \in \mathscr{L}} \left( \Phi_r \otimes I_n \right)\bar x_r,
\end{gathered}
\label{eq: observer containment error vector}
\end{equation}

\begin{lemma}[\cite{zuo2017output}]
\label{le: Phi_{nu} positive definite}
Given Assumption \ref{ass: leader follower}, $\sum_{r \in \mathscr{L}} {\Phi_r} $ is non-singular and positive-definite.
\end{lemma}

Subsequently, we integrate the attack-resilience with collision-free safety as follows.

\begin{lemma}
\label{le: specific CBF lemma}
Given the set \( S_{S_{ij}} \) and the CBF \( h_{S_{ij}}(x_i, x_j) \), if the control input \( u_i \) satisfies:
\begin{equation}
\begin{gathered}
\inf_{u_i \in \mathcal{U}_i} \left[ L_{f_i} h_{S_{ij}} + L_{g_i} h_{S_{ij}} u_i \right] \leqslant -\alpha(h_{S_{ij}}) \hfill\\
- 2(x_i - x_j)^\top A_j x_j - 2(x_i - x_j)^\top B_j u_j,
\end{gathered}
\label{eq: specific CBF condition}
\end{equation}
then the system ensures that \( \|x_i - x_j\| \geqslant d_s \) for all \( t \geqslant 0 \).
\end{lemma}

\noindent\textbf{\textit{Proof:}}
\label{prf: specific CBF lemma}
By definition, the specific CBF is given as:
\(
h_{S_{ij}}(x_i, x_j) = d_s^2 - \|x_i - x_j\|^2.
\) Taking the time derivative of \( h_{S_{ij}}(x_i, x_j) \) along the system dynamics, we have:
\(
\dot{h}_{S_{ij}}(x_i, x_j) = -\frac{d}{dt} \|x_i - x_j\|^2.
\) Expanding \( \|x_i - x_j\|^2 = (x_i - x_j)^\top (x_i - x_j) \), its time derivative is \(
\frac{d}{dt} \|x_i - x_j\|^2 = 2(x_i - x_j)^\top (\dot{x}_i - \dot{x}_j).
\)
Substituting this into \( \dot{h}_{S_{ij}}(x_i, x_j) \), we obtain:
\(
\dot{h}_{S_{ij}}(x_i, x_j) = -2(x_i - x_j)^\top (\dot{x}_i - \dot{x}_j).
\) From the system dynamics, we know: \(
\dot{x}_i = A_i x_i + B_i u_i, \quad \dot{x}_j = A_j x_j + B_j u_j.
\)
Substituting \( \dot{x}_i \) and \( \dot{x}_j \) into the expression for \( \dot{h}_{S_{ij}}(x_i, x_j) \), we get:
\(
\dot{h}_{S_{ij}}(x_i, x_j) = -2(x_i - x_j)^\top \left[ (A_i x_i + B_i u_i) - (A_j x_j + B_j u_j) \right].
\) Using the Lie derivative notation, this can be written as:
\(
\dot{h}_{S_{ij}}(x_i, x_j) = L_{f_i} h_{S_{ij}} + L_{g_i} h_{S_{ij}} u_i + 2(x_i - x_j)^\top A_j x_j + 2(x_i - x_j)^\top B_j u_j,
\)
where, \(L_{f_i} h_{S_{ij}} = -2(x_i - x_j)^\top A_i x_i, L_{g_i} h_{S_{ij}} = -2(x_i - x_j)^\top B_i\). Based on \cite{wang2017safety}, to ensure forward invariance of \( S_{S_{ij}} \), we impose the condition:
\[
\dot{h}_{S_{ij}}(x_i, x_j) \leqslant -\alpha(h_{S_{ij}}).
\]
Substituting the expanded form of \( \dot{h}_{S_{ij}}(x_i, x_j) \), we obtain:
\(
L_{f_i} h_{S_{ij}} + L_{g_i} h_{S_{ij}} u_i\leqslant -\alpha(h_{S_{ij}}) - 2(x_i - x_j)^\top A_j x_j - 2(x_i - x_j)^\top B_j u_j.
\) This inequality ensures that the safe set \( S_{S_{ij}} \) is forward invariant. Thus, if the control input \( u_i \) satisfies the optimization criterion:

\(
\inf_{u_i \in \mathcal{U}_i} \left[ L_{f_i} h_{S_{ij}} + L_{g_i} h_{S_{ij}} u_i \right] \leqslant -\alpha(h_{S_{ij}}) - 2(x_i - x_j)^\top A_j x_j - 2(x_i - x_j)^\top B_j u_j,
\)
then $\|x_i - x_j\| \geqslant d_s$ is guaranteed for all  $t \geqslant 0.\hfill\blacksquare$

\begin{figure}[!h]
\centering
\includegraphics[width=1.7in]{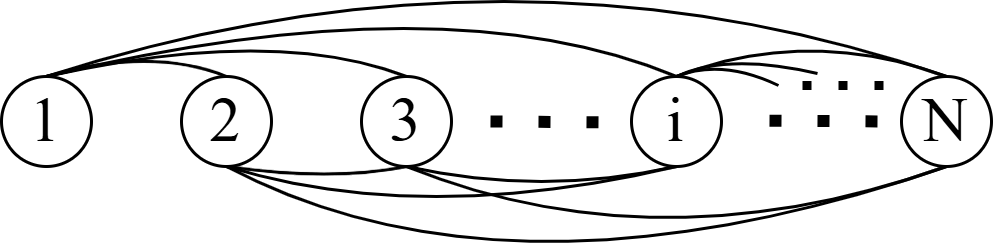}
\captionsetup{justification=centering}
\caption{Pairs.}
\label{fig: pairs}
\end{figure}

For control affine systems, CBFs lead to linear constraints on the control inputs $u_i$ that can be enforced online. To ensure safety constraints and prevent collisions, we design the SAAR control input \( u_i \) by solving the following QP. This optimization minimizes the difference between the actual SAAR control and the corrupted attack-resilient control input $\bar u_i$, while enforcing the constraints:
\begin{equation}
\min_{u_i} \| u_i - \bar u_i \|^2,
\label{eq: optimization problem to get u_i}
\end{equation}
subject to the constraints:
\begin{equation}
\begin{gathered}
\inf_{u_i \in \mathcal{U}_i} \left[ L_{f_i} h_{S_{ij}} + L_{g_i} h_{S_{ij}} u_i \right] \leqslant -\delta_{ij}h_{S_{ij}} \hfill\\
- 2(x_i - x_j)^\top A_j x_j - 2(x_i - x_j)^\top B_j u_j,
\end{gathered}
\label{eq: the safety constraints}
\end{equation}
where \(1 \leqslant i < j < N\), which ensures the pairwise distances are defined for all agent pairs $(i,j)$ and each pair is counted exactly once as shown in Fig~\ref{fig: pairs}. $\delta_{ij}$ is the parameters which influences the intensity of the safety constraints. We select $\alpha(h_{S_{ij}}) = \delta_{ij}h_{S_{ij}}$\cite{garg2020distributed}. Let $\Delta u_i=u_i - \bar u_i$, which is obtained by solving a convex QP with linear constraints over the feasible set \( \mathcal{U}_i \) (which is assumed to be compact or otherwise bounded in practice). Therefore, \(\Delta u_i\) is bounded. The efficient solution of the QP problem enables the algorithm to be implemented in real-time \cite{borrmann2015control}. Collision avoidance among agents is enforced through a sequential backward-constrained optimization scheme. Starting from the highest-indexed agent pair, the solution for each agent~$i$ is computed using constrained optimization based on results from previously solved pairs~$(i,j)$, where~$j>i$. For example, if the total number of followers $N=4$, the resulting pairs are $(1,2)$, $(1,3)$, $(1,4)$, $(2,3)$, $(2,4)$, and $(3,4)$, fully enumerating all distinct pairwise interactions. In the updating algorithm, inputs for agents are sequentially updated by first addressing pairs involving the highest-indexed agents. Specifically, the input of the highest-indexed agent remains unmodified, i.e., $u_4 = \bar u_4$, the input for agent 3 is updated based on the $(3,4)$ pair. Then, using the updated information for agents 3 and 4, the input for agent 2 is updated through pairs $(2,3)$ and $(2,4)$. Finally, inputs for agent 1 are updated simultaneously using the pairs $(1,2)$, $(1,3)$, and $(1,4)$, incorporating previously updated inputs. An alternative approach from \cite{wang2017safety} with reduced neighborhoods on a disk graph provides a scalable solution with less computational cost.

\begin{figure}[!h]
\centering
\includegraphics[width=3.65in]{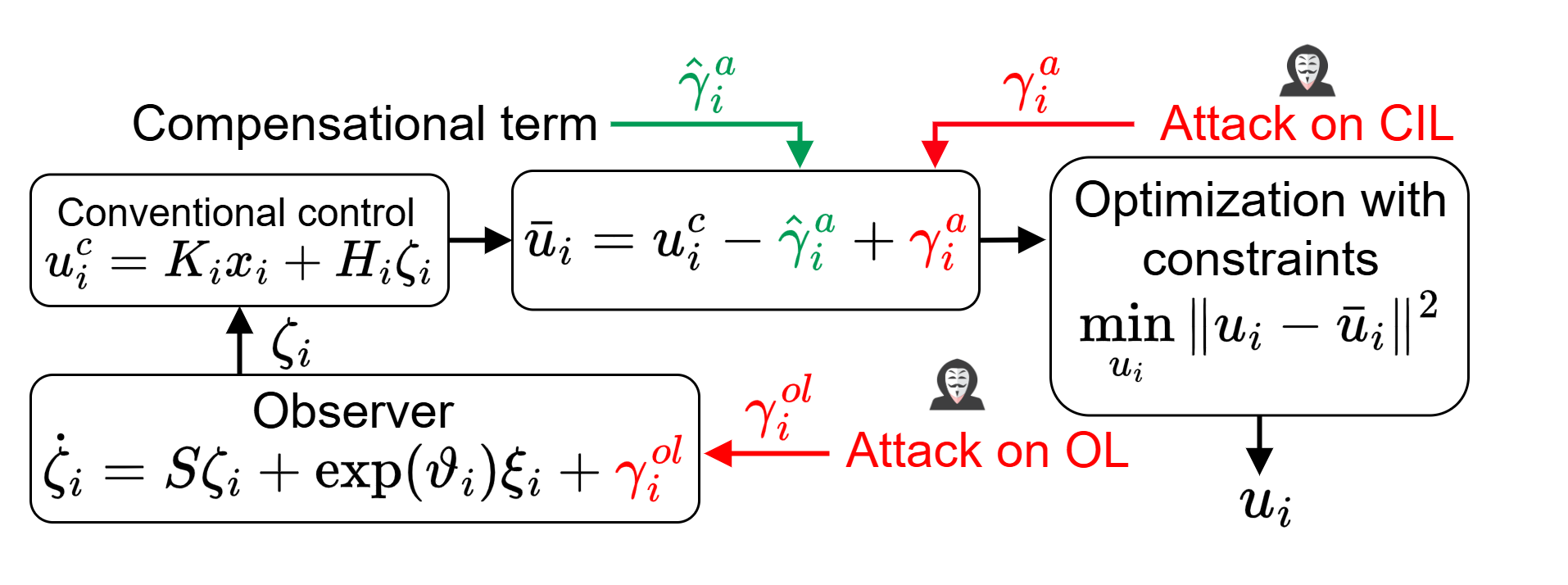}
\captionsetup{justification=centering}
\caption{Block diagram of the SAAR control system.}
\label{fig: safety diagram}
\end{figure}

The proposed SAAR control system is shown in Fig.~\ref{fig: safety diagram}.

\vspace{-1mm}
\begin{theorem}
\label{thm: controller convergence}
Given Assumptions \ref{ass: leader follower}- \ref{ass: attacks}, considering the heterogeneous MAS composed of \eqref{eq: follower dynamics}-\eqref{eq: leader dynamics} in the presence of EU-FDI attacks on both CIL and OL, the SAAR control problem is solved by designing the fully-distributed safe and resilient controller consisting of \eqref{eq: single agent observer dynamics}-\eqref{eq: adaptively tuned parameter} and \eqref{eq: optimization problem to get u_i}-\eqref{eq: the safety constraints}.
\end{theorem}
\noindent\textbf{\textit{Proof:}}
The global form of \eqref{eq: containment information} is 
\begin{equation}
\begin{gathered}
\xi= -\sum_{\nu \in \mathscr{L}} (\Phi_{\nu} \otimes I_N) \big( \zeta - \bar x_r\big)
= -\sum_{\nu \in \mathscr{L}} (\Phi_{\nu} \otimes I_N)\Delta_o ,\hfill
\end{gathered}
\label{eq: global xi}
\end{equation}
where $\xi= {[ {\xi_1^\top,...,\xi_N^\top} ]^\top}$. Based on Lemma \ref{le: Phi_{nu} positive definite}, $\sum_{\nu \in \mathscr{L}} (\Phi_{\nu} \otimes I_n)$ is nonsingular. To prove $\Delta_o$ is UUB is equivalent to proving that $\xi$ is UUB. The global form of $\dot \zeta_i$ in \eqref{eq: single agent observer dynamics} is
\begin{equation}
\begin{gathered}
\dot \zeta = (I_N\otimes S)\zeta + \operatorname{diag}(\exp(\vartheta_i))\xi + \gamma^{ol},\hfill
\end{gathered}
\label{eq: global form of zeta_i dot}
\end{equation}
where $\gamma^{ol}= {[ {{\gamma^{ol}_1}^\top,...,{\gamma^{ol}_N}^\top} ]^\top}$. The time derivative of $\xi$ in \eqref{eq: global xi} is
\begin{equation}
\begin{gathered}
\dot \xi= -\sum\nolimits_{\nu \in \mathscr{L}} (\Phi_{\nu} \otimes I_n) \big((I_N\otimes S)\zeta + \hfill\\
 \big(\operatorname{diag}(\exp(\vartheta_i))\otimes{I_n}\big)\xi + \gamma^{ol} - (I_N\otimes S)\bar x_r\big)\hfill\\
=(I_N\otimes S)\xi-\sum_{r \in \mathscr{L}} (\Phi_{r} \otimes I_n)\big(\operatorname{diag}(\exp(\vartheta_i))\otimes{I_n}\big)\xi\hfill\\
-\sum\nolimits_{r \in \mathscr{L}} (\Phi_{r} \otimes I_n)\gamma^{ol}.\hfill
\end{gathered}
\label{eq: global xi dot}
\end{equation}
We consider the following Lyapunov function candidate
\begin{equation}
V^{'} = \tfrac{1}{2}\sum\nolimits_{i \in \mathscr{F}} \xi_i^\top\xi _i \exp(\vartheta_i).
\label{eq: Lyapunov function}
\end{equation}

The time derivative of $V^{'}$ along the trajectory of \eqref{eq: global xi dot} is 
\begin{equation}
\scalebox{0.85}{$
\begin{gathered}
  \dot V^{'} = \sum\nolimits_{i \in \mathscr{F}} \big(\xi_i^\top{\dot \xi_i} \exp(\vartheta_i) + \frac{1}{2}\xi^T_i\xi_i\exp(\vartheta_i)\dot \vartheta_i\big)\hfill \\
  = \xi^\top\operatorname{diag}\big(\exp(\vartheta_i)\otimes I_N\big)\dot\xi + \frac{1}{2}\xi^T_i\big(\operatorname{diag}(\exp(\vartheta_i)\dot \vartheta_i)\otimes I_N\big)\xi \hfill\\
  = \xi^\top\operatorname{diag}\big(\exp(\vartheta_i)\otimes I_N\big)
  \bigg((I_N\otimes S)\xi - \sum_{r \in \mathscr{L}} (\Phi_{r} \otimes I_N)\hfill\\
  \times\big(\operatorname{diag}(\exp(\vartheta_i))\otimes{I_N}\big)\xi
  - \sum_{r \in \mathscr{L}} (\Phi_{r} \otimes I_N)\gamma^{ol}\bigg) + \frac{1}{2}\xi^\top\hfill\\
  \times\big(\operatorname{diag}(\dot \vartheta_i)\otimes I_N\big)\big(\operatorname{diag}(\exp(\vartheta_i))
  \otimes I_N\big)\xi\hfill\\
  \leqslant \sigma_{\max}(S)\|\big(\operatorname{diag}(\exp(\vartheta_i))\otimes{I_N}\big)\xi\|\|\xi\|- \sigma_{\min}\big(\sum_{r \in \mathscr{L}}\Phi_{r}\big)\hfill\\\times\|\big(\operatorname{diag}(\exp(\vartheta_i))\otimes{I_N}\big)\xi\|^2  + \sigma_{\max}\big(\sum_{r \in \mathscr{L}}\Phi_{r}\big)\hfill\\
  \times \|\big(\operatorname{diag}(\exp(\vartheta_i))\otimes{I_N}\big)\xi\|\|\gamma^{ol}\| + \frac{1}{2}\max_i(\dot \vartheta_i)\hfill\\
  \times\|\big(\operatorname{diag}(\exp(\vartheta_i))\otimes{I_N}\big)\xi\|\|\xi\|\hfill\\
  = - \sigma_{\min}\big(\sum_{r \in \mathscr{L}}\Phi_{r}\big)\|\big(\operatorname{diag}(\exp(\vartheta_i))\otimes{I_N}\big)\xi\|\hfill\\
  \times\bigg(\|\big(\operatorname{diag}(\exp(\vartheta_i))\otimes{I_N}\big)\xi\| - \sigma_{\max}(S)\hfill\\
  /\sigma_{\min}\big(\sum_{r \in \mathscr{L}}\Phi_{r}\big)\|\xi\| - \sigma_{\max}\big(\sum_{r \in \mathscr{L}}\Phi_{r}\big)/\sigma_{\min}\big(\sum_{r \in \mathscr{L}}\Phi_{r}\big)\hfill\\
  \times\|\gamma^{ol}\| - \frac{1}{2}\max_i(\dot \vartheta_i)/\sigma_{\min}\big(\sum_{r \in \mathscr{L}}\Phi_{r}\big)\|\xi\|\bigg).\hfill\\
\end{gathered}
$}
\label{eq: time derivative of the Lyapunov function}
\end{equation}

Denote $\phi_a = \sigma_{\max}(S)
  /\sigma_{\min}\big(\sum_{r \in \mathscr{L}}\Phi_{r}\big)$ and $\phi_b = \sigma_{\max}\big(\sum_{r \in \mathscr{L}}\Phi_{r}\big)/\sigma_{\min}\big(\sum_{r \in \mathscr{L}}\Phi_{r}\big)$, which are both positive constants. For $\dot V^{'} \leqslant 0$, we need
\begin{equation}
\begin{gathered}
\|\big(\operatorname{diag}(\exp(\vartheta_i))\otimes{I_N}\big)\xi\| - \phi_a\|\xi\| - \phi_b\|\gamma^{ol}\|\hfill\\
- \frac{1}{2}\max_i(\dot \vartheta_i)/\sigma_{\min}\big(\sum_{r \in \mathscr{L}}\Phi_{r}\big)\|\xi\|\geqslant 0. \hfill
\end{gathered}
\label{eq: critical part of dot V'}
\end{equation}
A sufficient condition to guarantee \eqref{eq: critical part of dot V'} is 
\begin{equation}
\begin{gathered}
\big(\exp(\vartheta_i) - \phi_a - \frac{1}{2}\max_i(\dot \vartheta_i)/\sigma_{\min}\big(\sum_{r \in \mathscr{L}}\Phi_{r}\big)\big)\|\xi_i\|\hfill\\
\geqslant\phi_b\|\gamma^{ol}_i\|.\hfill
\end{gathered}
\label{eq: the sufficient condition of time derivative of the Lyapunov function ge to 0}
\end{equation}
A sufficient condition to guarantee \eqref{eq: the sufficient condition of time derivative of the Lyapunov function ge to 0} is $\|\xi_i\|\geqslant\phi_b$ and $\exp(\vartheta_i) - \phi_a- 1/2\max_i(\dot \vartheta_i)/\sigma_{\min}(\sum_{r \in \mathscr{L}}\Phi_{r})\geqslant\|\gamma^{ol}_i\|$. From Assumption \ref{ass: attacks}, $\|\gamma^{ol}_i(t)\| \leqslant \exp(\kappa^{ol}_it)$, to prove that $\exp(\vartheta_i) - \phi_a - 1/2\max_i(\dot \vartheta_i)/\sigma_{\min}(\sum_{r \in \mathscr{L}}\Phi_{r})\geqslant\|\gamma^{ol}_i\|$, we need to prove that $\exp(\vartheta_i) - \phi_a - 1/2\max_i(\dot \vartheta_i)/\sigma_{\min}(\sum_{r \in \mathscr{L}}\Phi_{r})\geqslant\exp(\kappa^{ol}_it)$. Based on \eqref{eq: adaptive gain theta_i dynamics}, when $\|\xi_i\|>\max\{\sqrt{\kappa^{ol}_i/q_i},\phi_b\}$, which guarantees the exponential growth of $\exp(\vartheta_i)$ dominates all other terms, $\exists t_1$, such that $\forall t > t_1$, $\exp(\vartheta_i) - \phi_a - 1/2\max_i(\dot \vartheta_i)/\sigma_{\min}(\sum_{r \in \mathscr{L}}\Phi_{r})\geqslant\exp(\kappa^{ol}_it)$. Hence, we obtain $\forall t > t_1$,
\begin{equation}
\dot V^{'} \leqslant 0,\:\forall \|\xi_i\|>\max\{\sqrt{\kappa^{ol}_i/q_i},\phi_b\}.
\label{eq: dot Lyapunov function candidate le 0}
\end{equation}
By LaSalle’s
invariance principle \cite{lasalle1960some}, $\xi_i$ is UUB. Therefore, $\Delta_o$ is UUB.
Next, we prove that follower-observer tracking error $\varepsilon_i$ is UUB.
From \eqref{eq: follower dynamics}, \eqref{eq: solution for tracking}, \eqref{eq: single agent observer dynamics}, \eqref{eq: control input} and \eqref{eq: H_i equation}, we obtain the time derivative of \eqref{eq: follower-observer error vector} as
\begin{equation}
\begin{gathered}
  {{\dot \varepsilon }_i} = {{\dot x}_i} - {\dot \zeta }_i \hfill \\
   = {A_i}{x_i} + {B_i}{K_i}{x_i} + {B_i}{H_i}{\zeta _i} - {B_i}{\hat \gamma^a_i} \hfill \\
   + {B_i}{\gamma^a_i} - \left( A_i + B_i \Pi_i \right) \zeta_i - \operatorname{exp}(\vartheta_i) \zeta_i + \Delta u_i - \gamma_i^{ol} \hfill \\
= \left( A_i + B_i K_i \right) \varepsilon_i - B_i \hat{\gamma}_i^a - B_i \gamma_i^a - \operatorname{exp}(\vartheta_i) \xi_i - \gamma_i^{ol} + \Delta u_i. \hfill
\end{gathered}
\label{eq: epsilon dot}
\end{equation}
From the above proof, we confirmed ${\xi _i}$ is UUB. Considering Assumption \ref{ass: eig of S}, \eqref{eq: global xi} and \eqref{eq: global form of zeta_i dot}, we obtain that $\beta_i=\operatorname{exp}(\vartheta_i) \xi_i - \gamma_i^{ol}$ is bounded. Let ${{\bar A}_i} = {A_i} + {B_i}{K_i}$ and ${\bar Q_i}={Q_i} + K_i^\top{U_i}{K_i}$. Note that ${\bar Q_i}$ is positive-definite. From \eqref{eq: Riccati equation}, $P_i$ is symmetric positive-definite. Consider the following Lyapunov function candidate
\begin{equation}
V_i = \varepsilon_i^TP_i\varepsilon_i,
\label{eq34}
\end{equation}
and its time derivative is given by
\begin{equation}
\begin{gathered}
\dot V_i = 2\varepsilon^T_iP_i\dot\varepsilon_i\hfill\\
= 2 \varepsilon_i^TP_i\left(\bar{A}_i \varepsilon_i+B_i \gamma^a_i-B_i \hat\gamma^a_i - \beta_i + \Delta u_i\right)\hfill\\
\leqslant-\sigma_{\min }\left(\bar{Q}_i\right)\left\|\varepsilon_i\right\|^2+2\left(\varepsilon_i^TP_i B_i \gamma^a_i-\varepsilon_i^TP_i B_i \hat\gamma^a_i\right)\hfill\\
- 2 \varepsilon_i^TP_i\beta_i + 2 \varepsilon_i^TP_i\Delta u_i\hfill\\
\leqslant-\sigma_{\min }\left(\bar{Q}_i\right)\left\|\varepsilon_i\right\|^2+2\left(\varepsilon_i^TP_i B_i \gamma^a_i-\varepsilon_i^TP_i B_i \hat\gamma^a_i\right)\hfill\\
+2 \sigma_{\max }\left(P_i\right)\left\|\varepsilon_i\right\|\left\|\beta_i\right\| +2 \sigma_{\max }\left(P_i\right)\left\|\varepsilon_i\right\|\left\|\Delta u_i\right\|.\hfill\\
\end{gathered}
\label{eq35}
\end{equation}
Using \eqref{eq: compensational signal} to obtain
\begin{equation}
\begin{gathered}
\varepsilon_i^\top P_i B_i \gamma^a_i-\varepsilon_i^\top P_i B_i \hat\gamma^a_i \hfill\\
= \varepsilon_i^\top P_i B_i \gamma^a_i-\frac{\left\|\varepsilon_i^\top P_i B_i\right\|^2}{\left\|\varepsilon_i^\top P_i B_i\right\|+\exp \left(-c_i t^2\right)} \exp \left(\hat{\rho}_i\right)\hfill\\
\leqslant \left\|\varepsilon_i^\top P_i B_i\right\|\left\|\gamma^a_i\right\|-\frac{\left\|\varepsilon_i^\top P_i B_i\right\|^2}{\left\|\varepsilon_i^\top P_i B_i\right\|+\exp \left(-c_i t^2\right)} \exp \left(\hat{\rho}_i\right)\hfill\\
=\left\|\varepsilon_i^\top P_i B_i\right\|\big(\left\|\varepsilon_i^\top P_i B_i\right\|\left\|\gamma^a_i\right\|+\exp(-c_i t^2)\left\|\gamma^a_i\right\|\hfill\\  
-\left\|\varepsilon_i^\top P_i B_i\right\|\exp\left(\hat{\rho}_i\right)\big)/\big(\left\|\varepsilon_i^\top P_i B_i\right\|+\exp(-c_i t^2)\big).\hfill
\end{gathered}
\label{eq36}
\end{equation}
To prove that $\varepsilon_i^\top P_i B_i \gamma^a_i-\varepsilon_i^\top P_i B_i \hat\gamma^a_i \leqslant 0$, we need to prove that $\left\|\varepsilon_i^\top P_i B_i\right\|\left\|\gamma^a_i\right\| + \exp(-c_i t^2)\left\|\gamma^a_i\right\| -\left\|\varepsilon_i^\top P_i B_i\right\|\exp\left(\hat{\rho}_i\right)\leqslant 0$. Define $\upsilon_i = \kappa^a_i/\sigma_{\operatorname{min}}(P_iB_i)$, $\omega_i = 2 \sigma_{\max }\left(P_i\right)\big(\left\|\beta_i\right\| + \|\Delta u_i\|\big)/\sigma_{\min }\left(\bar{Q}_i\right)$. Then, define the compact sets $\Upsilon_i\,\triangleq\,\{\|\varepsilon_i\|\leqslant\upsilon_i\}$ and $\Omega_i\,\triangleq\,\{\|\varepsilon_i\|\leqslant\omega_i\}$. Considering Assumption \ref{ass: attacks}, we obtain that $\exp(-c_i t^2)\left\|\gamma^a_i\right\|\rightarrow 0$. Hence, outside the compact set $\Upsilon_i=\ \{\|\varepsilon_i\|\leqslant\upsilon_i\}$, $\exists t_1$, such that $\varepsilon_i^\top P_i B_i \gamma^a_i-\varepsilon_i^\top P_i B_i \hat\gamma^a_i \leqslant 0$, $\forall t \geqslant t_1$; outside the compact set $\Omega_i$, $-\sigma_{\min }\left(\bar{Q}_i\right)\left\|\varepsilon_i\right\|^2+2 \sigma_{\max }\left(P_i\right)\left\|\varepsilon_i\right\|\left\|\beta_i\right\| \leqslant 0$. Therefore, combining \eqref{eq35}, \eqref{eq36} and \eqref{eq: epsilon dot}, we obtain, outside the compact set $\Upsilon_i\cup\Omega_i$, $\forall t \geqslant t_1$,
\begin{equation}
\dot V_i\leqslant 0.
\label{eq: Vi dot le 0}   
\end{equation}
Hence, by the LaSalle’s
invariance principle, $\varepsilon_i$ is UUB. We conclude that $\Delta_o$ and $\varepsilon_i$ are UUB, consequently, $e_c=\varepsilon+\Delta_o$ is UUB. Furthermore, based on Lemma~\ref{le: specific CBF lemma}, under the satisfaction of safety constraints \eqref{eq: the safety constraints}, we obtain that $h_{S_{ij}} \leqslant 0$. That is $\|x_i - x_j\| \geqslant d_s$. This completes the proof.$\hfill\blacksquare$

\section{Numerical Simulations}

\label{sec:NumericalSimulations}
\begin{figure}[!h]
\centering
\includegraphics[width=1.1in]{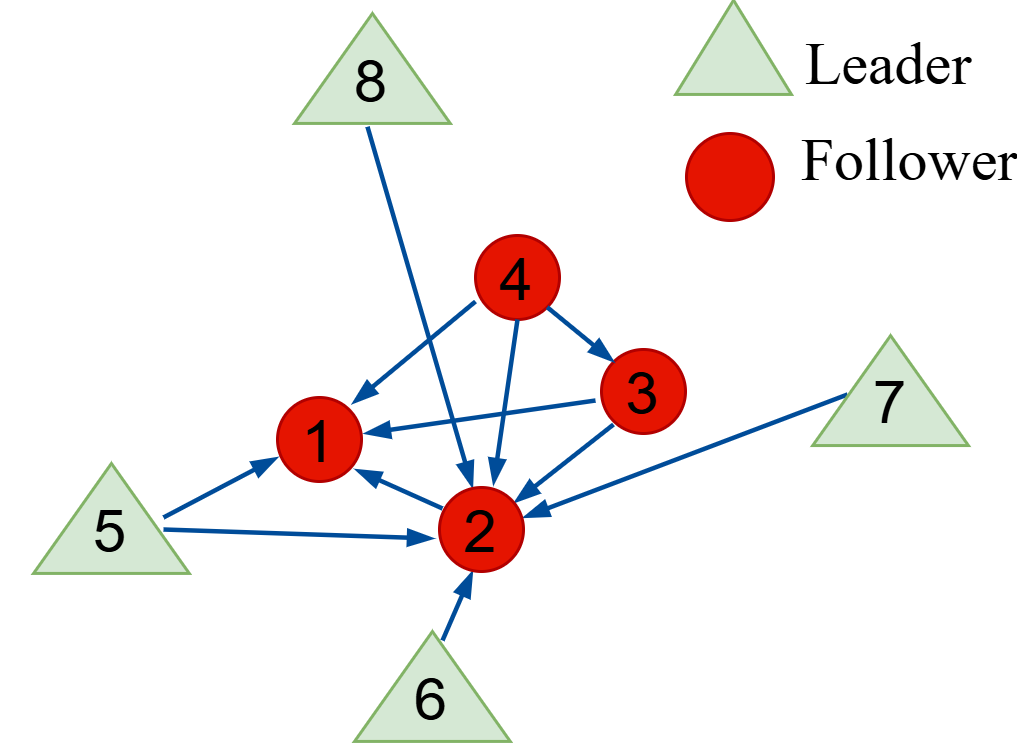}
\captionsetup{justification=centering}
\caption{Communication topology.}
\label{fig: Communication topology and the attacks}
\end{figure}

We validate the proposed SAAR control strategies through numerical simulations of a heterogeneous autonomous MAS under EU-FDI attacks on both CIL and OL. Fig.~\ref{fig: Communication topology and the attacks} illustrates the communication topology among the agents. As seen, the autonomous MAS consists of four followers (agents 1 to 4) and four leaders (agents 5 to 8), with the following dynamics:
\[
\scalebox{0.7}{$
\begin{aligned}
A_1 &= \begin{bmatrix}-2 & 1 & 0\\0 & -3 & 1\\0.5 & 0 & -1\end{bmatrix},\quad
B_1 = \begin{bmatrix}1 & 0 & 0\\0 & 1 & 0\\0 & 0 & 1\end{bmatrix},\quad
A_2 = \begin{bmatrix}-1 & 0 & 0.5\\0 & -2 & 1\\0.5 & 0 & -0.5\end{bmatrix},\\[4pt]
B_2 &= \begin{bmatrix}0.5 & 1 & 0\\1 & 0.5 & 0\\0 & 0 & 1\end{bmatrix},\quad
A_3 = \begin{bmatrix}-1 & 1 & 0\\0 & -3 & 1\\0 & 0.5 & -1\end{bmatrix},\quad
B_3 = \begin{bmatrix}1 & 0 & 0\\0 & 1 & 0\\0 & 0 & 1\end{bmatrix},\\[4pt]
A_4 &= \begin{bmatrix}-1 & 0.5 & 0\\0.5 & -1.5 & 0.5\\-0.5 & 0 & -2\end{bmatrix},\quad
B_4 = \begin{bmatrix}1 & 0 & 0\\0 & 1 & 0\\0 & 0 & 1\end{bmatrix},\quad
S = \begin{bmatrix}0 & -2 & 1\\2 & 0 & 1\\-1 & -1 & 0\end{bmatrix}.
\end{aligned}
$}
\]

We consider the following EU-FDI attack signals on the CIL and OL: 
\[
\scalebox{0.7}{$
\begin{array}{cc}
    \gamma^{a}_{1} = 
    \begin{bmatrix}
        2.5 e^{0.07t} \\
        1.5 e^{0.04t} \\
        -6.6 e^{0.08t}
    \end{bmatrix},
    \gamma^{ol}_{1} = 
    \begin{bmatrix}
        -1.2 e^{0.10t} \\
        1.5 e^{0.17t} \\
        2.7 e^{0.15t}
    \end{bmatrix},
    
    \gamma^{a}_{2} = 
    \begin{bmatrix}
        2.3 e^{0.05t} \\
        -4.7 e^{0.05t} \\
        11.5 e^{0.04t}
    \end{bmatrix},
    \gamma^{ol}_{2} = 
    \begin{bmatrix}
        3.3 e^{0.06t} \\
        -2.2 e^{0.15t} \\
        -1.7 e^{0.12t}
    \end{bmatrix}, \\[1.5em]
    
    \gamma^{a}_{3} = 
    \begin{bmatrix}
        3.6 e^{0.10t} \\
        -4.7 e^{0.09t} \\
        -10.2 e^{0.06t}
    \end{bmatrix},
    \gamma^{ol}_{3} = 
    \begin{bmatrix}
        2.8 e^{0.14t} \\
        -5.0 e^{0.04t} \\
        -1.8 e^{0.08t}
    \end{bmatrix},
    
    \gamma^{a}_{4} = 
    \begin{bmatrix}
        -2.9 e^{0.09t} \\
        5.2 e^{0.06t} \\
        -7.7 e^{0.07t}
    \end{bmatrix},
    \gamma^{ol}_{4} = 
    \begin{bmatrix}
        -5.2 e^{0.04t} \\
        2.4 e^{0.13t} \\
        -2.1 e^{0.14t}
    \end{bmatrix}.
\end{array}
$}
\]
Select $U_{1,2,3,4}=I_2$, and $Q_{1,2,3,4}=3I_2$. The controller gain matrices $K_i$ and $H_i$ are found by solving \eqref{eq: K_i equation}-\eqref{eq: Riccati equation}. 

\begin{figure}[!h]
\centering
\includegraphics[width=2.7in]{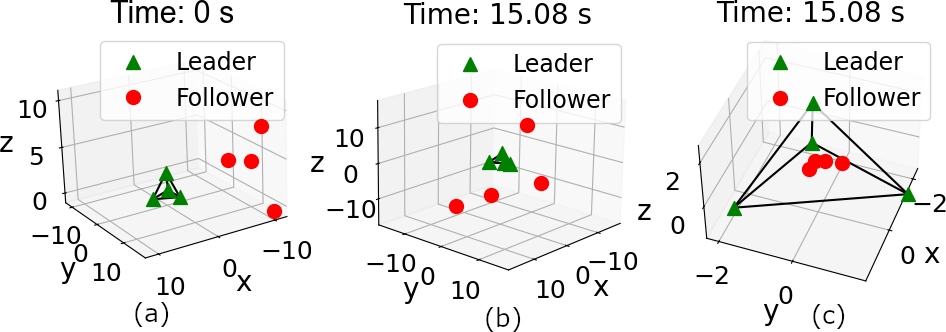}
\caption{Leader-follower 3D motion evolution: (a) $0\,\operatorname{s}$ snapshot. (b) $15.08\,\operatorname{s}$ snapshot (after attack initiation) using the standard control. (c) $15.08\,\operatorname{s}$ snapshot (after attack initiation) using the SAAR control.}
\label{fig: Leader-follower motion evolution snapshots.}
\end{figure}

We choose $d_s = 0.3$ for collision avoidance. The performance of the proposed SAAR control strategies, considering these safety constraints, is demonstrated through a set of comparative simulation results contrasted with non-SAAR standard control protocols \cite{zuo2017output}. The three sub-figures in Fig.~\ref{fig: Leader-follower motion evolution snapshots.} illustrate the initial 3D positions of the agents and compare the system responses under standard control and SAAR control following the attacks. Both EU-FDI attacks on the OL and CIL are initiated at $t=3\,\operatorname{s}$. Fig.~\ref{fig: Leader-follower motion evolution snapshots.}(a) depicts the initial positions of the agents, while Fig.~\ref{fig: Leader-follower motion evolution snapshots.}(b) presents the agents' 3D positions at $t=15.08\,\operatorname{s}$ under the standard control approach, where the containment objective is not achieved after attack initiation. In contrast, Fig.~\ref{fig: Leader-follower motion evolution snapshots.}(c) displays the agents' 3D positions at $t=15.08\,\operatorname{s}$ under the SAAR control approach, demonstrating that the UUB containment objective is preserved despite the attacks. 

\begin{figure}[!h]
\centering
\includegraphics[width=2.9in]{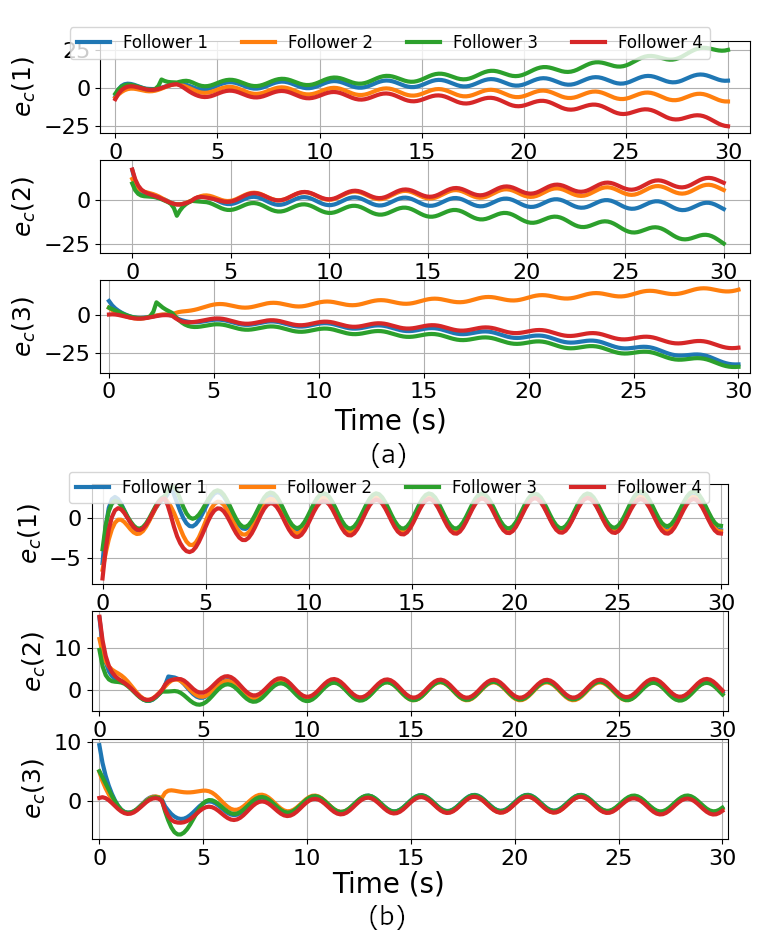}
\captionsetup{justification=centering}
\caption{Containment errors: (a) using conventional control (b) using SAAR control.}
\label{fig: containment errors}
\end{figure}

Fig.~\ref{fig: containment errors}(a) illustrates the evolution of the containment error \( e_c \) under conventional control. The dimensional components of \( e_c \) diverge, indicating a failure in achieving containment. In contrast, Fig.~\ref{fig: containment errors}(b) depicts the system response under SAAR control under attack injection. The SAAR controller quickly compensates after the attacks are initiated, causing the errors to stabilize and enter a steady state. This behavior demonstrates the resilience of the SAAR controller in maintaining UUB containment performance by effectively mitigating the impact of EU-FDI attacks on both CIL and OL.

\begin{figure}[!h]
\centering
\includegraphics[width=3.4in]{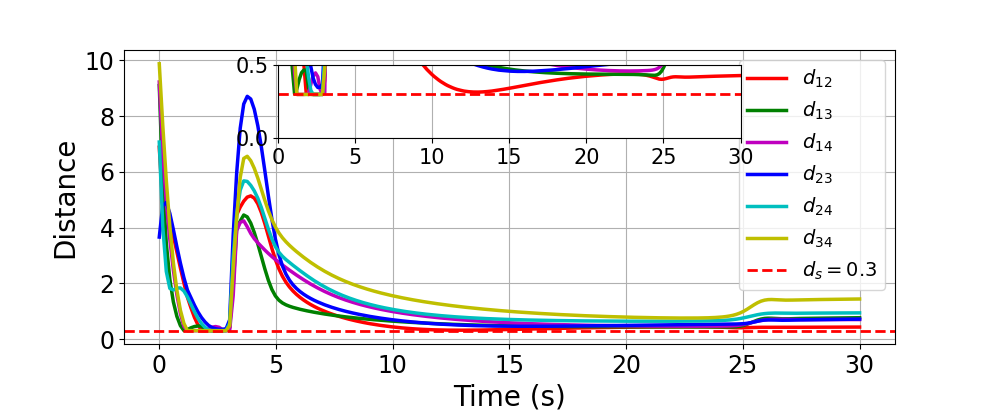}
\captionsetup{justification=centering}
\caption{Pairwise distance.}
\label{fig: pairwise distance}
\end{figure}

Fig.~\ref{fig: pairwise distance} depicts the evolution of the distances between followers, denoted as \( d_{12}, d_{13},d_{14},d_{23},d_{24}, \) and \( d_{34} \), over time, which shows the show safe distances among followers are maintained. Fig.~\ref{fig: containment errors}(b) and Fig.~\ref{fig: pairwise distance} show that the system successfully maintains safety and achieves the UUB containment under EU-FDI attacks on both OL and CIL, validating the effectiveness of the proposed SAAR defense strategies.

\section{Conclusion}
This letter has formulated and solved the SAAR control problem for heterogeneous MASs to mitigate EU-FDI attacks on both OL and CIL while ensuring collision-free behaviors. An attack-resilient OL and a novel compensational signal on the CIL have been designed to address the EU-FDI attacks. A SAAR controller has then been developed by formulating and solving a QP problem integrating CBF certified collision-free safety constraints. Rigorous Lyapunov stability analysis and 3D simulations have demonstrated the SAAR controller’s effectiveness in maintaining both safety and resilience in practical autonomous MAS environments, where leader agents span safe regions to guide heterogeneous follower agents while ensuring collision avoidance.

\bibliographystyle{IEEEtran}

\bibliography{References}

\end{document}